\documentclass[4apaper]{aipproc}
\layoutstyle{8x11single}
\begin{document}
\author{J. Delabrouille, J. Kaplan\\
on behalf of the Planck HFI consortium}
{address={PCC, Coll{\`e}ge de France, 11 
place Marcelin Berthelot, 75231 Paris cedex 05}}
\title{Measuring CMB polarisation with the Planck HFI}
\begin{abstract}
    
The Planck High Frequency Instrument (HFI) is the most sensitive 
instrument currently being built for the measurement of Cosmic 
Microwave Background anisotropies.  In addition to unprecendented 
sensitivity to CMB temperature fluctuations, the HFI has 
polarisation-sensitive detectors in 3 frequency channels (143, 217 and 
353 GHz), which will constrain full-sky polarised 
emission of the CMB and foregrounds at these frequencies.  The sensitivity 
of the instrument will allow a clear detection of CMB polarisation 
signals and should yield a precise measurement of its power spectrum 
at all angular scales between $\ell = 50$ and $\ell = 1000$, as well 
as constraints on the polarised emission at larger scales where a 
polarised signal from inflationary gravity waves or from reionisation 
is expected in many cosmological scenarios.
\end{abstract}
\maketitle

\section{Introduction}

The COBE-DMR \cite{dmr-sci1}, Boomerang 
\cite{boomerang-sci1,boomerang-sci2}, DASI \cite{dasi-sci1,dasi-sci2} 
and Maxima \cite{maxima-sci1,maxima-sci2,maxima-sci3} experiments 
together now have yielded strong constraints on the Cosmic Microwave 
Background (CMB) anisotropy power spectrum \cite{jaffe-2001}, and in 
particular a convincing detection of the first three acoustic peaks, 
providing compelling evidence that indeed the primordial fluctuations 
were produced during an inflationary phase.  The next great challenge 
in the the study of the statistical properties of the CMB is to 
measure the polarisation signal, both that due to acoustic 
oscillations in causally connected regions of the Universe before 
decoupling and, even more challenging, the primordial polarisation 
spectrum due to gravity waves generated during inflation.  Measurement 
of the correlated spectrum of polarisation and temperature at 
sub-degree scales will provide yet another test of the basic acoustic 
oscillation scenario and of the global paradigm, while measurement (or 
absence of detection) of polarisation signals due to tensor modes 
could strongly constrain inflationary models, as well as yield direct 
observational evidence for the existence of primordial gravity waves 
\cite{zaldarriaga-1997,kamionkowski-1997a,seljak-1997,kamionkowski-1997b}.

This has been widely recognised by the CMB community, and a large 
number of experiments dedicated to detecting the CMB polarisation are 
currently in operation or being planned.  In this paper, we review the 
overall design of the Planck High Frequency Instrument (HFI) as a 
polarisation sensitive instrument, and discuss its capabilities in 
terms of measuring CMB polarisation.

\section{The Planck High Frequency Instrument}

The Planck mission, to be launched by ESA in spring 2007, is the third 
generation satellite dedicated to observing CMB anisotropies.  The DMR 
instrument on COBE yielded the first detection of CMB anisotropies 
on large angular scales.  The MAP mission, launched by NASA in June 
2001, will provide a measurement of anisotropies at 15 arcminute 
resolution with a sensitivity of a few tens of $\mu$K per resolution 
element.  The Planck mission will measure CMB anisotropies with a 
sensitivity yet an order of magnitude better, a few $\mu$K over 7 
arcminute resolution pixels.

The HFI is one of the two instruments on board Planck.  It is a 
48--detector instrument, using bolometers cooled to 100 mK by an 
open--cycle spatial dilution fridge.  Its 48 detectors are distributed 
into 6 frequency channels ranging from 100 to 850 GHz.  Originally 
designed as a temperature-sensitive instrument only, it has been 
modified for polarisation sensitivity in three of its frequency 
channels.  In the current (final) design, half of the detectors of the 
Planck HFI are polarisation sensitive (see table \ref{tab:1}).

The other instrument on the Planck spacecraft, the Low Frequency 
Instrument (LFI), polarisation-sensitive as well, is described by 
Villa et al. in these proceedings \cite{villa-here}.

\begin{center}

\begin{table}[h]
\begin{tabular}{l|cccccc}
    Central Frequency (GHz)        & 100 & 143 & 217 & 353 & 545 & 857  
    \\
    \hline
    Beam size (arcmin)             & 9.2 & 7.1 & 5.0 & 5.0 & 5.0 & 
    5.0  \\
    N det. unpolarised             &  4  &  4  &  4  &  4  &  4  &  
    4   \\
    N det. polarised               &  -  &  8  &  8  &  
    8  &  -  &  -   \\
    I sensitivity ($\mu$K/K)       & 2.2 & 2.4 & 3.8 & 15  & 80  & 
    8000 \\
    U and Q sensitivity ($\mu$K/K) &  -  & 4.8 & 7.6 & 
    30  &  -  &  -   \\
    Flux sensitivity (mJy)         & 9.0 & 12.6 & 9.4 & 20 & 46  & 
    52   \\
\end{tabular} 
\caption{Summary of Planck HFI main characteristics. Sensitivities 
for intensity and polarisation are given per square pixel with a 
beam size side, for a 1 year mission.}
\label{tab:1} 
\end{table}

\end{center}

\subsection{Polarisation-sensitive bolometers}

\begin{figure}[h]  
     {\includegraphics[scale=0.5]{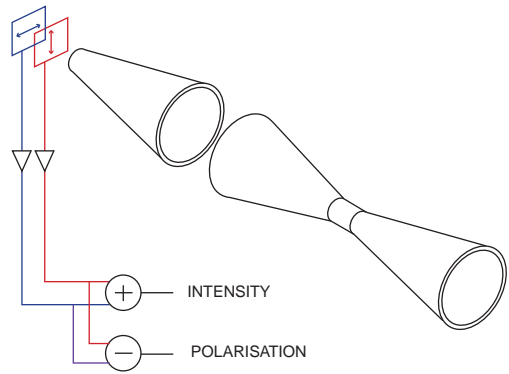}}    
     \caption{Schematic view of the Planck HFI polarisation dispositive 
     using Polarisation Sensitive Bolometers.} 
     \label{fig:1}  
     \end{figure}

The Planck HFI Polarisation Sensitive Bolometers (PSB) are 
built so that two sensors using absorbers made with parallel wires 
\cite{bock}, coupled to orthogonal polarisation modes, are located in 
the same integration cavity.  The two sensitive devices coupled in 
this way share the same optics (horns, filters, telescope), but are 
different detectors (with each its own coupling efficiency, time 
constant, sensitivity). Each of them is read out by its own read-out 
electronics chain, as illustrated in Figure \ref{fig:1}.

With this design, the optical coupling to the sky should be identical 
for the two elements of the PSB (same beam shape, same response as a 
function of the wavelength), except for the sensitivity to orthogonal 
polarisation directions.  Numerical simulations of the instrumental 
beam show that differences in beam shapes between the two polarisation 
directions of a single PSB are at negligible level (typically one per 
cent effect or less) \cite{yurchenko,fosalba}.  Noise levels, time 
constants of the detection chain (or more generally its impulse 
response), however, may differ slightly between the two elements, 
although at the manufacturing level it is hoped to match as well as 
possible the time constants and noise levels at the optimum (minimum) 
value.

The amount of depolarisation achieved with this design has been 
measured in-lab to be less than a few percent \cite{maffei}, with a 
target of about 3 \%, which should negligibly impact the sensitivity 
to polarised signals.

Each PSB, by differencing its two data streams, directly provides a 
measurement of the $Q$ Stokes parameter ({\em i.e.} a pure 
polarisation signal) in its own reference system (on an axis system 
where the $x$ and $y$ coordinates are measured along the two 
orthogonal polarisation-sensitive directions).

\subsection{PSB layout}

As emphasized in \cite{couchot-1999}, at least three (linear) 
polarisation sensitive measurements with different orientations are 
needed to fully measure $I$, $Q$, and $U$ in a given direction.  For 
perfectly matched, uncorrelated noise in the data, the minimal error 
box volume, as well as uncorrelated noise between the measurements of these 
three Stokes parameters, are realised when the polarimeter 
orientations are evenly distributed between 0 and 2$\pi$.  Couchot et 
al.  \cite{couchot-1999} denote such configurations (using $n \geq 3$ 
evenly distributed polarimeters) {\em optimised configurations}.

\begin{figure}[h]  
     {\includegraphics[scale=0.5]{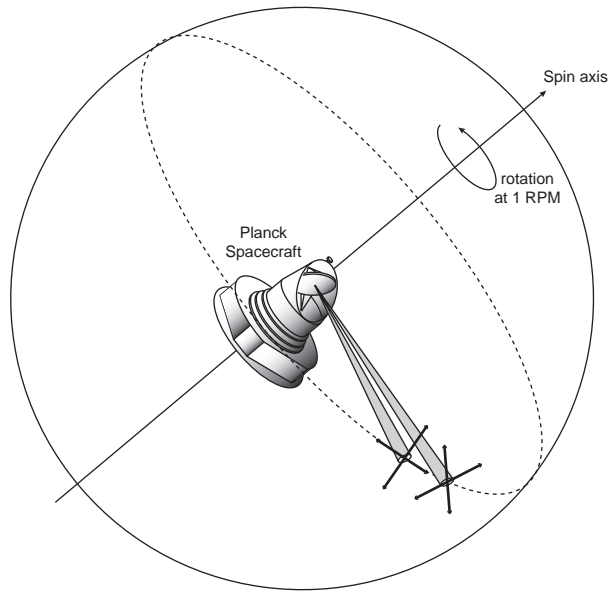}}    
     \caption{Polarisation measurement in the context of the Planck 
     scan-strategy. For each PSB-pair detector set, the 4 measurements 
     correspond to polarimeter orientations at 0, 45, 90 and 135 
     degrees with respect to the direction defined by the scanning.} 
     \label{fig:2}  
     \end{figure}

Then, a single PSB yielding two orthogonal polarisation measurements 
is not by itself sufficient to measure the three Stokes parameters 
$I$, $Q$, and $U$.  The layout of the detectors for the Planck HFI is 
such that every PSB has a companion oriented at 45 degrees, with 
relative locations in 
the focal plane such that during scanning, it measures 
polarisation on the same pointing trajectory a fraction of a second 
before or after the first one (see Figure \ref{fig:2}).  Such a pair 
of companion PSBs yields four timelines which correspond effectively, 
after rephasing the measurements to get co-extensive pointings, to a 
polarisation measurement with a 4-polarimeter {\em optimised 
configuration}.

Such a configuration has several advantages.  The matching of the 
orientations with the scanning direction ensures that one of the PSBs 
permits (by simple difference) to measure $Q$ and the second $U$ in 
the reference system where one axis is parallel to the scanning and the 
other one perpendicular.  The fact that two polarimeters share the 
same horn permits the best rejection of leakage of intensity $I$ 
signals into polarisation data, as discussed later.  In addition, 
there is in principle built-in redundancy, as if even one detector 
fails, the three remaining timelines are sufficient to obtain the three 
Stokes parameters of interest (even if they do not then constitute 
an {\em optimised configuration}).  This built-in redundancy permits 
an internal consistency check of the data (and thus of 
systematic effects) if all detectors work properly (in terms of 
sensitivity).

For each polarisation-sensitive frequency channel (at 143, 217 and 
353 GHz), there are two such companion PSB pairs, which provides yet 
another redundancy level, as well as better overall sensitivity.

\begin{figure}[h]  
     {\includegraphics[scale=0.5]{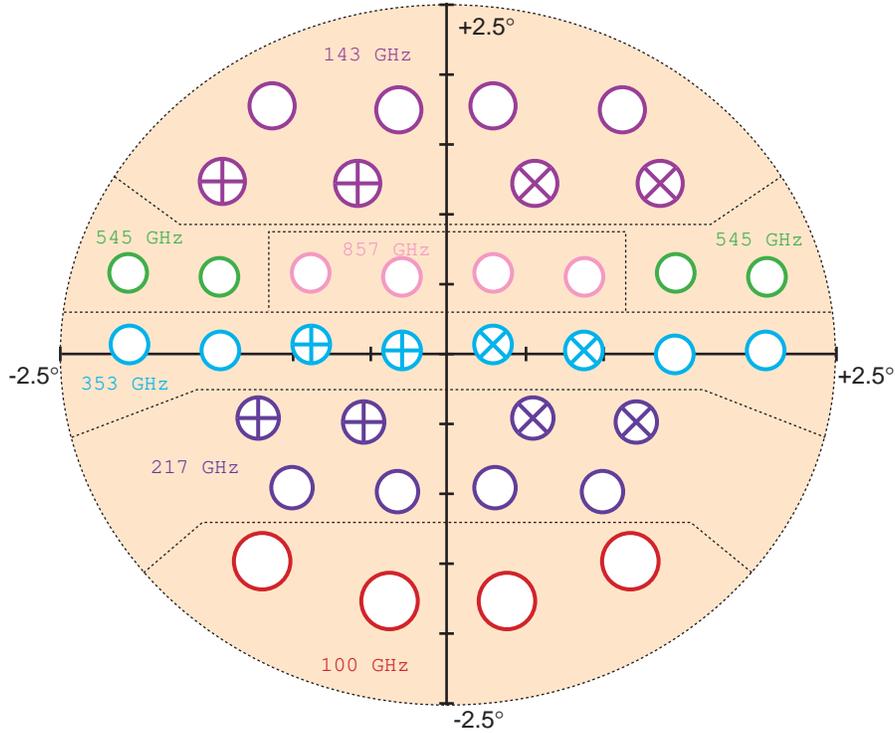}}   
     \caption{Planck HFI detector layout. In this figure, the 
     horizontal and vertical axis are parallel and perpendicular to 
     the line of scanning respectively. Polarisation bolometer fields 
     of view are represented with a cross inside, showing the 
     orientations of polarisation sensitivities.} 
     \label{fig:3}  
     \end{figure}

The HFI detector layout in the focal plane is shown in Figure \ref{fig:3}.

\section{Polarisation data processing}

Whereas a lot of experience has been already gathered by the 
scientific community in the analysis of CMB anisotropy data, no such 
know-how has yet been acquired for the specific measurement of 
polarisation.  Special effort has been undertaken in the HFI Data 
Processing Center (DPC) to develop data reduction tools specifically 
tailored and optimised for Planck HFI polarisation data.  In this 
section, we discuss (non exhaustively) some of the issues adressed by 
the Planck HFI DPC so far, although 
part of the discussion applies (and we believe can be useful) to other 
instruments as well.

\subsection{Modelling the measurement}

There are several particularities to the measurement of polarisation, 
as compared to temperature, that will require the implementation of 
specific data processing software for the processing of such data.  
The specifics of polarisation signals come from two origins:

\begin{enumerate}
    \item particularities of the signals we want to measure and which 
    impact the data processing: very weak signals, poorly known 
    galactic foregrounds and systematics;
    \item specifics of the instrumental set-up for polarisation measurements.
\end{enumerate}

Neglecting for the moment instrumental imperfections, each polarimeter 
of the Planck HFI, sensitive only to one linear polarisation, measures 
in each sample a linear combination of $I$, $Q$ and $U$ (integrated 
over the detector beam):
\begin{equation}
    d(\alpha) = \frac{1}{2} \, [I + Q \cos 2 \alpha + U \sin 2 \alpha]
\end{equation}
where $\alpha$ is the angle between the polarimeter orientation and 
the $x$-axis used for defining the Stokes parameters. 

\subsection{Combining data samples}

At least three independent samples, with different angles $\alpha$ but 
coextensive beam pointings, are required to measure $I$, $Q$ and $U$.  
For an ideal measurement with two PSBs, assuming perfectly coincident 
beam pointings, the data from the four timelines are given by:

\begin{equation}
    \left[ \begin{array}{c} d_{1} \\ d_{2} \\ d_{3} \\ d_{4} 
    \end{array} \right] =
    \left[ \begin{array}{c} d(0) \\ d(\pi/2) \\ d(\pi/4) \\ d(3\pi/4)
    \end{array} \right] = \frac{1}{2}
    \left[ \begin{array}{ccc} 
         1 & 1 & 0 \\ 
         1 & -1 & 0 \\ 
         1 & 0 & 1 \\ 
         1 & 0 & -1
    \end{array} \right] 
    \left[ \begin{array}{c} I \\ Q \\ U \end{array} \right] \, + \,
    \left[ \begin{array}{c} n_{1} \\ n_{2} \\ n_{3} \\ n_{4} \end{array} \right]  
\end{equation}    
where $n_{1},\ldots n_{4}$ are noise terms, and $I$, $Q$, $U$ denote 
{\em the same} beam-integrated Stokes parameters on the sky in the 
pointing direction.  This equation can be recast in the matrix form:
\begin{equation}
    d = AS + n
\end{equation}    
where $d$ is the data, $A$ a $4 \times 3$ matrix, $S$ the 
three-element Stokes-parameter vector, and $n$ the noise.

With this ideal setup, for well balanced, uncorrelated noise, the best 
Stokes parameters estimates in the PSB-pair frame are obtained as 

\begin{equation}
    \begin{array}{ll}
    I & = \frac{1}{4} \times (d_{1}+d_{2}+d_{3}+d_{4}) \\
    Q & = \frac{1}{2} \times (d_{1}-d_{2})  \\
    U & = \frac{1}{2} \times (d_{3}-d_{4})
    \end{array}
\end{equation}    

For unbalanced and/or correlated noise, the optimal least square solution is
\begin{equation}
    \widetilde S = [A^tN^{-1}A]^{-1} \, A^tN^{-1}d
\end{equation}    
where $N$ is the noise covariance matrix, $N_{ij} = \langle 
n_{i}n_{j}\rangle$.  

\subsection{Monitoring systematic effects}

So far, we have not said anything about unavoidable noise correlations 
along the time streams, we have just considered the measurement at a 
single given pointing direction.  In addition, all four detectors of a 
PSB pair have been assumed to have coextensive beams, which is 
actually not the case.  The actual data processing for polarised 
map-making has to take into account these imperfections of the 
measurement.  We concentrate on two major issues, which are low 
frequency drifts and pointing mismatch, and show how these 
imperfections impact polarised data processing for the Planck HFI.

\subsubsection{Low frequency drifts}

Low frequency drifts, due both to $1/f$ noise in the detection 
chain and to thermal fluctuations of payload elements detected by the 
sensors, are expected to be present in the detector timelines.  This 
will be the case for PSB timelines as much as for unpolarised 
bolometers timelines.  For each polarimeter $i$, writing explicitely 
the summmation over pixels to avoid confusion on other repeated 
indices, the data stream can be modelled as:
\begin{equation}
    d_{it} = \sum_{p}M_{itp}[I_{p} + Q_{p} \cos 2 \alpha_{itp} + U_{itp} \sin 2 
    \alpha_{itp}] + n_{it}.
    \label{eq:model1}
\end{equation}
where $d_{it}$ is the data of polarimeter $i$ at time $t$, 
$(I_{p},Q_{p},U_{p})$ are the Stokes parameters in pixel $p$, 
$M_{itp}$ is the pointing matrix for polarimeter $i$, telling how much 
pixel $p$ contributes to the signal of polarimeter $i$ at time $t$.  
$\alpha_{itp}$ is the angle, in pixel $p$, between the co-polar 
direction of polarimeter $i$ at time $t$ and the reference direction 
(e.g. parallel and perpendicular to longitude and latitude lines), and 
$n_{it}$ is the noise timestream for detector $i$.  Usually for 
temperature measurements, the pointing matrix is modelled as a sparse 
matrix containing only one non-vanishing element per line, which just 
tells which pixel of the sky is pointed at at time $t$.

Reordering the four timelines of a PSB pair $d_{it}$ into a single 
data vector $d$, the four noise streams into one single noise vector 
$n$, and the three maps of Stokes parameters $I_{p}$, $Q_{p}$ and 
$U_{p}$ into a single vector $S$, we can recast the four equations in
eq. \ref{eq:model1} (one for each detector) into one single linear 
equation:
\begin{equation}
    d=MS+n.
\end{equation}
Denoting as $N$ the noise autocorrelation (which now encompasses both 
correlations {\em along} timelines and correlations {\em between} 
timelines), the best estimates for the Stokes parameters, $\widetilde 
S$, can in principle be optained by a Global Least Square (GLS) 
inversion of the linear equation, $\widetilde S = [M^tN^{-1}M]^{-1} \, 
M^tN^{-1}d$. This however is a formidable task for Planck HFI, even 
for temperature maps, because of the size of the system. It is even 
more so with polarisation data.

A first-order map-making algorithm which approximates the GLS solution 
has been developped in \cite{revenu-2000} to construct polarisation 
maps from all timelines of an {\em optimised configuration} in the 
context of the Planck HFI. The method simplifies the resolution of the 
linear system under the assumption that noise correlations on 
timescales less than 60 seconds are negligible.  Although a useful 
step towards a global solution, this simplified solution (which 
does not take into account imperfections of the model as it assumes a 
perfect knowledge of the pointing) can not be used for Planck in 
its present implementation.  For polarisation mapping indeed, special 
care has to be taken in the implementation of the inversion of the 
linear system, as discussed in the next paragraphs.


\subsubsection{Pointing mismatch}

In the formal GLS solution, it is implicitly assumed that the matrix $M$, 
which encompasses both the pointing of the detectors and the mixing of 
the Stokes parameters, is known and perfectly describes the actual 
detector pointing.  The fact that this is not exactly true complicates 
the processing of polarisation data in a somewhat tricky way.

\begin{figure}[h]  
     {\includegraphics[scale=0.4]{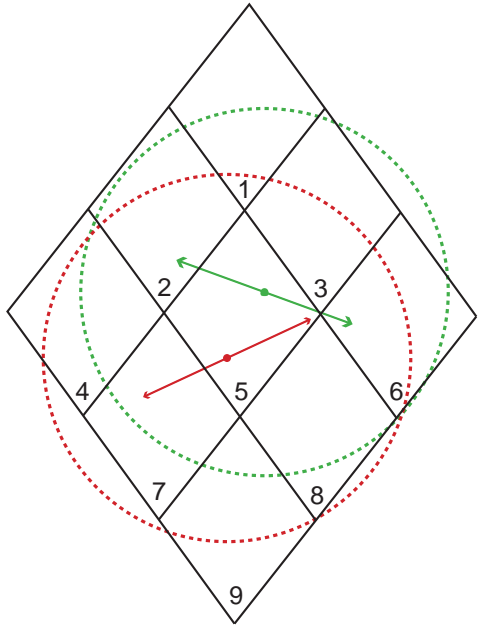}}      
     \caption{Illustration of the effect of single pixel pointing 
     approximation or of pointing mismatch due to improper knowledge 
     of the exact pointing.  Here, in pixel 5, two polarisation 
     measurements with different orientations are not pointed at the 
     exact same place.  Even if the beams are much bigger than the 
     pixel, as should be for proper sampling, the difference between 
     the two measurements includes a term proportional to the gradient 
     of the beam-smoothed temperature map, which may be larger than 
     the actual average polarisation in that pixel.  For a careless 
     implementation of a GLS map-making solution directly from 
     polarimeter data streams, this may become a ``killer--effect'' 
     for polarisation measurements.}
     \label{fig:4}  
     \end{figure}

Let us assume that a number $m$ of polarised data samples, all 
corresponding to a pointing towards the same given pixel $p$, are to 
be combined to recover a best estimate of $I$, $Q$ and $U$ at that 
point.  For a pixel size $\theta_{p}$, the measurements are actually 
not pointing all to the exact same place, as illustrated in Figure 
\ref{fig:4}.  Any gradient of the temperature through the pixel will 
generate a difference in the readout of two polarimeters which do not 
integrate signal from perfectly co-extensive beams.  For mismatched 
pointings, if the two polarimeters do not have the same orientation, a 
temperature gradient term will leak into the estimated polarisation 
(obtained through difference terms in the inversion of the linear 
system). This pointing error effect, even if essentially small or 
even negligible for temperature measurements, can nonetheless be 
important for polarisation because of the relative levels of $I$, $Q$ 
and $U$ on the sky.

For each pixel, the fake polarisation generated in this way will 
depend on the exact distribution of all pointings inside the pixel and 
on the distribution of corresponding polarimeter orientations.  
Relative pointings, therefore, are to be monitored very carefully for 
a polarisation-measuring experiment using single polarisers.

The naive model of the very sparse mixing matrix with one single 
non-vanishing element per line can be refined.  The exact pointing, 
however, has to be perfectly well known, which is not the case in any 
of the present instruments (and not the case for Planck).  The order 
of magnitude of the pointing reconstruction accuracy required for this 
effect to be smaller than the expected CMB polarisation for Planck is 
about 30 arcseconds.  For the effect to be completely negligible, a 
relative pointing reconstruction of typically 5 arcseconds or better 
is needed.  This requirement becomes less stringent in the limit of many 
measurements per pixel, as the effect of random uncorrelated errors in 
pointing reconstruction will tend to cancel out.

\subsubsection{Planck HFI polarised map-making}

The use of Polarisation Sensitive Bolometers (or of any setup in which 
two orthogonal polarimeters share the same optics) provides an elegant 
solution to the pointing reconstruction accuracy problem, as beams are 
in principle perfectly co-extensive.  Then, instead of implementing a 
GLS map-making on single polarimeter timelines, the map-making can be 
implemented on differences between two orthogonal polarimeter 
readouts.  If the transfer functions of the two polarimeters in each 
PSB (time constants and readout impulse response) are well matched 
(which can be done by numerical post-filtering the two timelines if 
not built in hardware) the temperature gradient leakage problem is 
solved.  This method is currently the baseline polarised map-making 
solution for nominal HFI performance (co-extensive PSB beams, nominal 
balanced noise levels, pointing reconstruction accuracies of a few 
tens of arcseconds or worse).

Additionnal discussion on polarisation systematics can be found 
in \cite{kaplan-here}.

\subsubsection{Polarised component separation}

Polarisation maps obtained with Planck in general, and the Planck HFI 
in particular, will contain at each frequency a mixture of the 
polarised emission of several astrophysical components.  Although the 
component separation can be made based on the same general principles 
as for temperature maps, the overall performance of the separation is 
still unclear.  First order methods assuming prior knowledge of 
the emission spectra and the spatial spectra have been developped 
\cite{bouchet-1999b} to generalise the Wiener separation method first 
implemented by Bouchet \& Gispert \cite{bouchet-1999a} and Tegmark \& 
Efstathiou \cite{tegmark-1996}.  However, as little prior information 
on the polarised emission of foregrounds is actually available, 
blind separation methods 
\cite{baccigalupi-2000,maino-2001,snoussi-2001} must be investigated 
in this particular context.

Note still that for the Planck HFI, contrarily to the temperature case 
where several astrophysical foregrounds are expected to have 
temperature contribution larger than, or of the order of, the noise 
level, foreground polarised emission is expected to be well below both 
the polarisation sensitivity and the CMB polarisation, making 
component separation less critical for CMB polarisation measurements 
than for temperature anisotropies mapping (fig. \ref{fig:5}).

\section{The sensitivity of the Planck HFI}

A detailed modelling of the performance of the HFI bolometers onboard 
Planck in the background conditions at the L2 Sun-Earth Lagrange point 
permits to predict the sensitivity of the HFI in each channel.  For 
this estimation, it is assumed that the observation time is evenly 
shared between all the sky pixels.  Corresponding estimated 
sensitivites on I, Q and U per resolution side square pixel, for all 
HFI frequency channels, are given in table \ref{tab:1}.

\begin{figure}[h]  
     \includegraphics[scale=0.45]{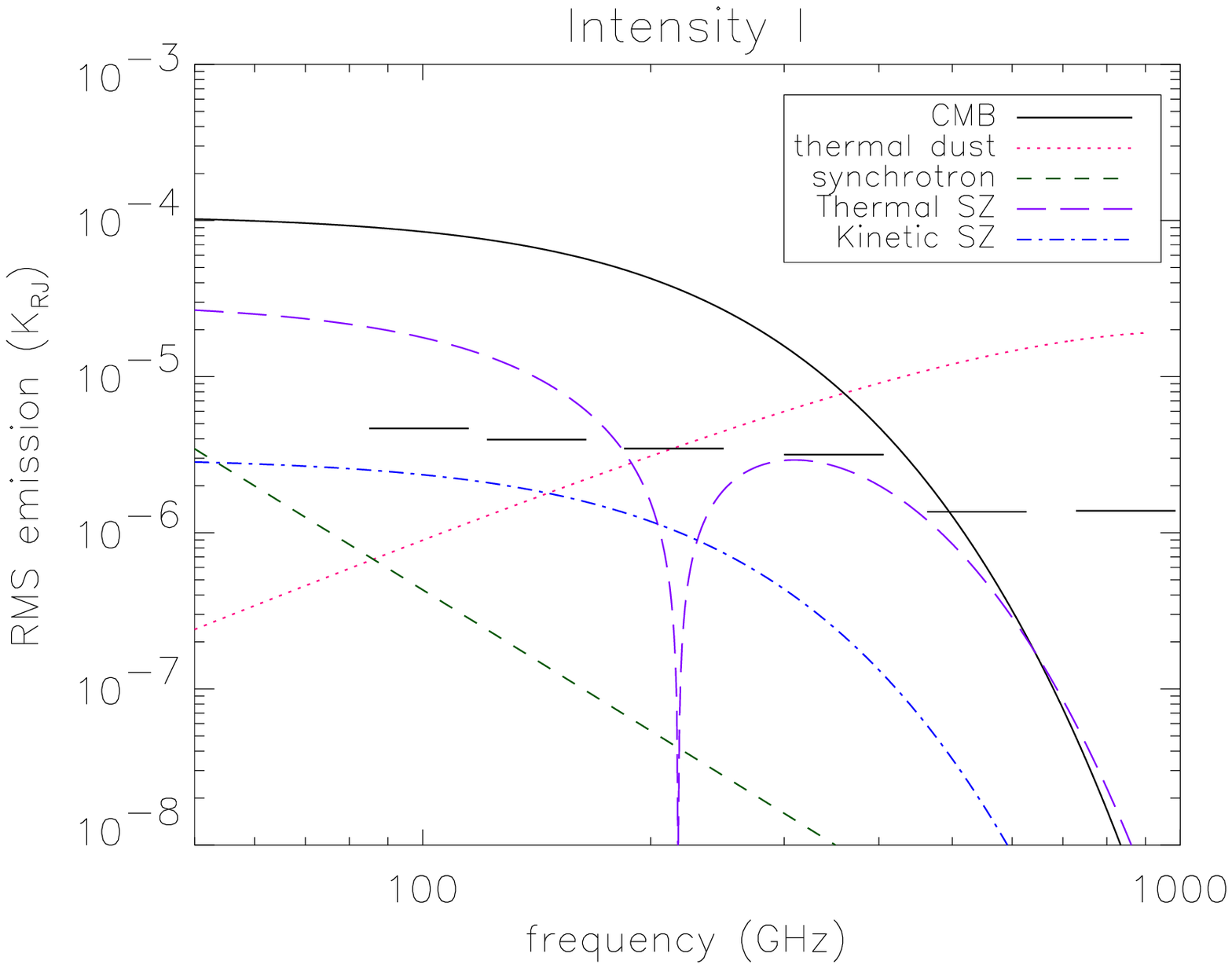} \hspace{.5cm}  
     \includegraphics[scale=0.45]{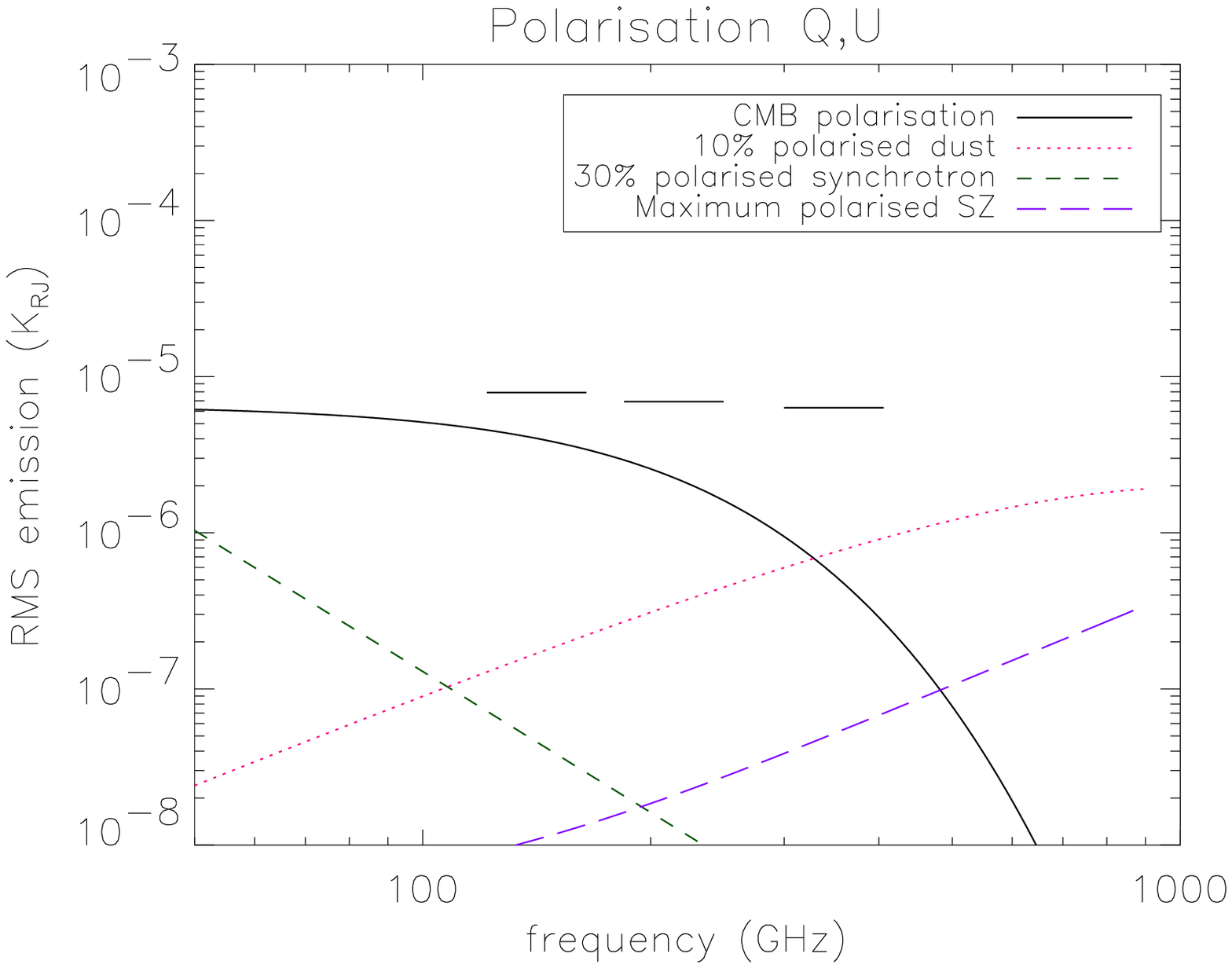} 
     \caption{RMS intensity (left) and polarised emission (right) of 
     various astrophysical components at high galactic latitude ($b 
     \simeq 70^\circ$).  Planck HFI Intensity and Polarisation 
     sensitivities per resolution element in all frequency bands 
     appear as horizontal bars on the plots.}
     \label{fig:5}  
     \end{figure}

\begin{figure}[h]  
     {\includegraphics[scale=0.45]{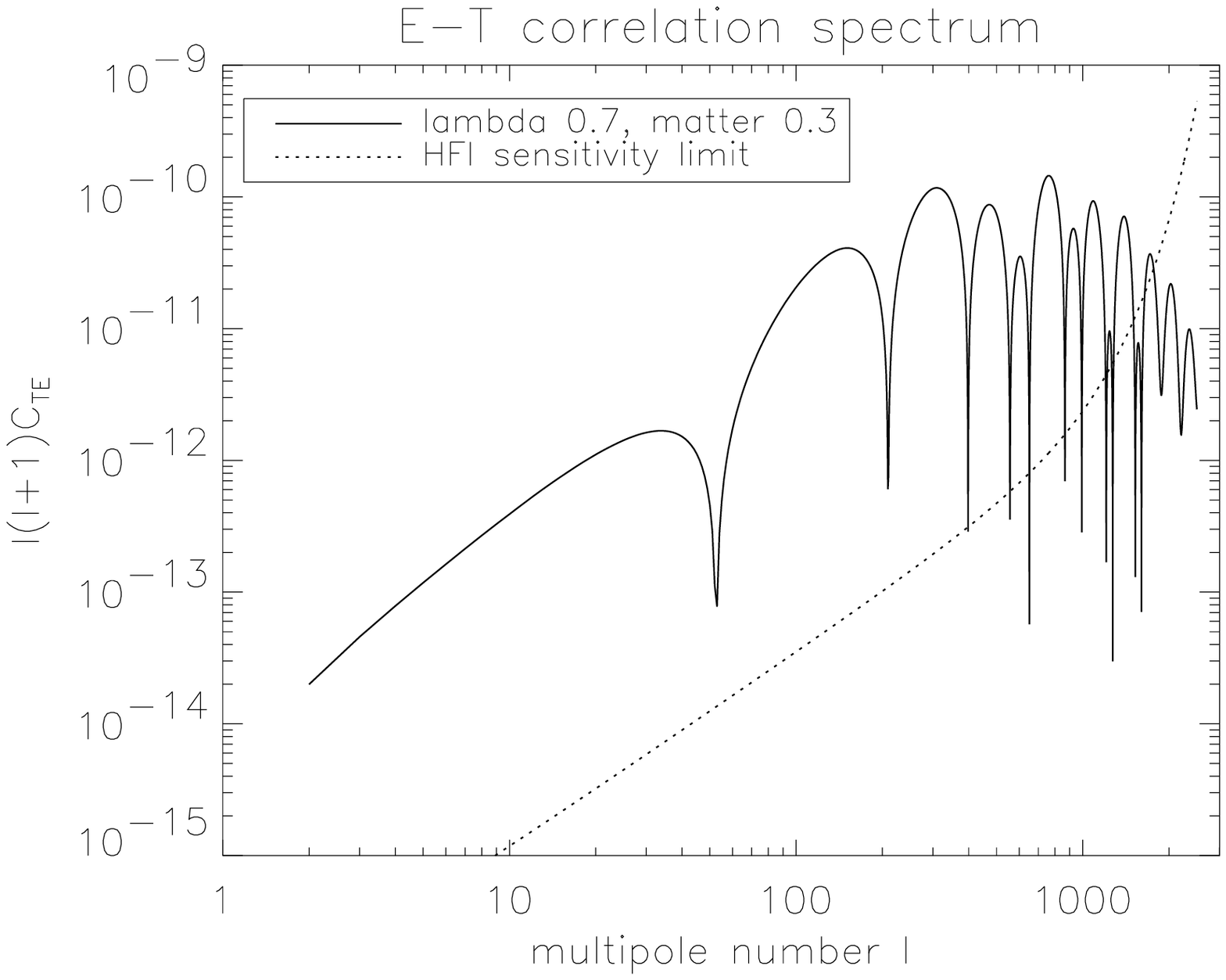}}\hspace{.5cm}   
     {\includegraphics[scale=0.45]{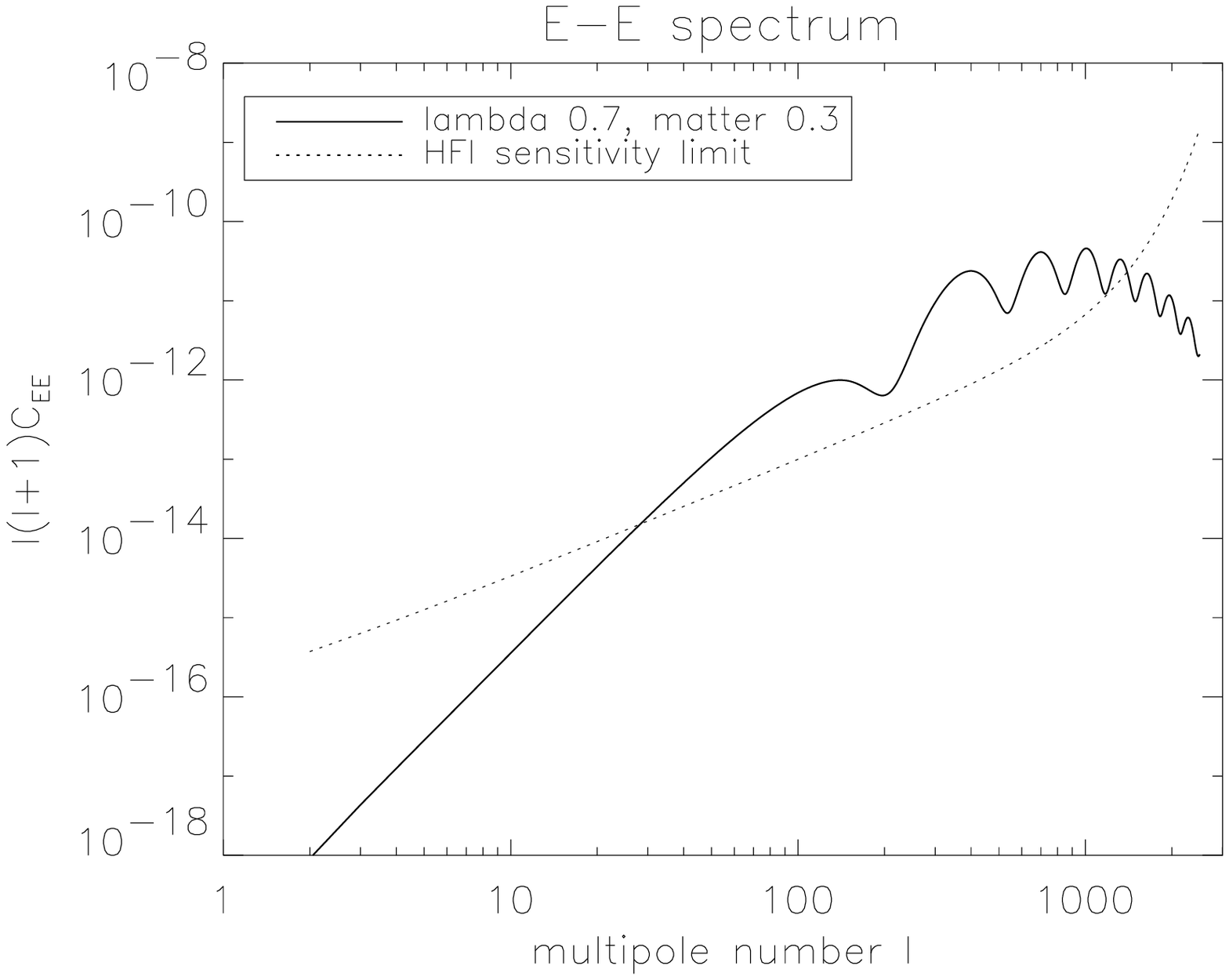}}    
     \caption{HFI sensitivity to $E-T$ cross correlation (left) and to 
     $E-E$ spectrum (right) for the 143 GHz HFI channel only.  Power 
     spectra are plotted for scalar modes only in a CDM-like 
     cosmological model with $\Omega_{b}=0.045$, $H_{0} = 65 \, {\rm 
     km/s/Mpc}$, $\Omega_{\Lambda}=0.7$ and $\Omega_{m}=0.3$, and have 
     been obtained using the CMBFAST software \cite{cmbfast}.}
     \label{fig:6}  
     \end{figure}

The sensitivity of the Planck HFI to intensity and polarisation is 
shown in Figure \ref{fig:5}.  Galactic foreground emission level 
estimates are shown for high galactic latitudes ($b \simeq 70^ \circ 
$).  Sensitivity levels per resolution element ($1 \sigma$) are shown 
as horizontal lines for six unpolarised frequency bands and three 
polarised frequency bands.  The polarisation sensitivity is not 
sufficient to detect CMB polarisation at the level of individual 
pixels, nor to map foreground polarised emission at high galactic 
latitudes.  Still, the power spectrum sensitivity on the 143 GHz 
channel alone (assuming the other channels are used as foreground and 
systematics monitors) is good enough to measure the $E-T$ and $E-E$ 
spectra quite well (Figure \ref{fig:6}).

Most interestingly, the sensitivity of the Planck HFI is at a level 
which should allow to put strong contrains on $B$ modes, which are 
expected in a large range of cosmological scenarios to be much larger 
at low $\ell$ values than the $E$ modes of Figure \ref{fig:6}, and 
thus within the detection reach of Planck (especially after combining 
the data from all HFI and LFI channels).

\section{Conclusion}

Originally planned and proposed solely as a CMB temperature anisotropy 
sensitive instrument, the Planck HFI has been revised to become as 
well a CMB polarisation sensitive instrument.  Unprecedented 
sensitivity at high resolution is achieved through the combination of 
the use of new polarisation sensitive bolometers, cooled to 100 mK on 
a spaceborne mission with low background, and of the selection of 
observing frequency bands where diffraction limits the resolution 
at the 5 to 7 arcminute level.  In addition, the selected 
frequency range is at the expected minimum of the polarised emission 
from galactic foregrounds and extragalactic compact sources.

A substantial effort is being made to understand the impact of all 
possible systematic instrumental effects throughout the detection 
process and the data reduction pipeline.  The minimisation of such 
systematic errors through a rigorous choice of the instrumental setup, 
careful on-ground testing, and continuing dedicated effort to the 
development and optimisation of data reduction methods for 
polarisation measurements with the Planck HFI, give us confidence that 
the instrument can meet its ambitious objectives.

\section{Acknowledgements}

We thank all the people who have contributed in a way or an other to 
the preparation of the measurement of polarisation with the Planck 
HFI. Special thanks to Yannick Giraud-H\'eraud, Jean-Michel Lamarre, 
Michel Piat, C\'ecile Renault and Cyrille Rosset for useful 
discussions and dedicated work.  Thanks also to Jim Bartlett for 
useful suggestions towards improving the manuscript, and to Radek 
Stompor for interesting discussions about polarised map-making.

\bibliographystyle{aipproc}

\end{document}